\documentstyle[12pt]{article}

\catcode`\@=11
\long\def\@makefntext#1{
\protect\noindent \hbox to 3.2pt {\hskip-.9pt
$^{{\ninerm\@thefnmark}}$\hfil}#1\hfill}		

\def\ZZ{Z\!\!\!Z}
\def\RR{I\!\!R}

 \def\@makefnmark{\hbox to 0pt{$^{\@thefnmark}$\hss}}  

\def\ps@myheadings{\let\@mkboth\@gobbletwo
\def\@oddhead{\hbox{}
\rightmark\hfil\ninerm\thepage}
\def\@oddfoot{}\def\@evenhead{\ninerm\thepage\hfil
\leftmark\hbox{}}\def\@evenfoot{}
\def\sectionmark##1{}\def\subsectionmark##1{}}

\newcounter{sectionc}\newcounter{subsectionc}\newcounter{subsubsectionc}
\renewcommand{\section}[1] {\vspace{0.6cm}\addtocounter{sectionc}{1}
\setcounter{subsectionc}{0}\setcounter{subsubsectionc}{0}\noindent
	{\bf\thesectionc. #1}\par\vspace{0.4cm}}
\renewcommand{\subsection}[1] {\vspace{0.6cm}\addtocounter{subsectionc}{1}
	\setcounter{subsubsectionc}{0}\noindent
	{\it\thesectionc.\thesubsectionc. #1}\par\vspace{0.4cm}}
\renewcommand{\subsubsection}[1]
{\vspace{0.6cm}\addtocounter{subsubsectionc}{1}
	\noindent {\rm\thesectionc.\thesubsectionc.\thesubsubsectionc.
	#1}\par\vspace{0.4cm}}

\newcounter{appendixc}
\newcounter{subappendixc}[appendixc]
\newcounter{subsubappendixc}[subappendixc]

\renewcommand{\appendix}[1] {\vspace{0.6cm}
        \refstepcounter{appendixc}
        \setcounter{figure}{0}
        \setcounter{table}{0}
        \setcounter{equation}{0}
        \renewcommand{\thefigure}{\Alph{appendixc}.\arabic{figure}}
        \renewcommand{\thetable}{\Alph{appendixc}.\arabic{table}}
        \renewcommand{\theappendixc}{\Alph{appendixc}}
        \renewcommand{\theequation}{\Alph{appendixc}.\arabic{equation}}
        \noindent{\bf Appendix \theappendixc #1}\par\vspace{0.4cm}}

\renewenvironment{thebibliography}[1]
	{\begin{list}{\arabic{enumi}.}
	{\usecounter{enumi}\setlength{\parsep}{0pt}
\setlength{\leftmargin 1.25cm}{\rightmargin 0pt}
	 \setlength{\itemsep}{0pt} \settowidth
	{\labelwidth}{#1.}\sloppy}}{\end{list}}

\newcounter{itemlistc}
\newcounter{romanlistc}
\newcounter{alphlistc}
\newcounter{arabiclistc}

\newcommand{\fcaption}[1]{
        \refstepcounter{figure}
        \setbox\@tempboxa = \hbox{\tenrm Fig.~\thefigure. #1}
        \ifdim \wd\@tempboxa > 6in
           {\begin{center}
        \parbox{6in}{\tenrm\baselineskip=12pt Fig.~\thefigure. #1}
            \end{center}}
        \else
             {\begin{center}
             {\tenrm Fig.~\thefigure. #1}
              \end{center}}
        \fi}

\newcommand{\tcaption}[1]{
        \refstepcounter{table}
        \setbox\@tempboxa = \hbox{\tenrm Table~\thetable. #1}
        \ifdim \wd\@tempboxa > 6in
           {\begin{center}
        \parbox{6in}{\tenrm\baselineskip=12pt Table~\thetable. #1}
            \end{center}}
        \else
             {\begin{center}
             {\tenrm Table~\thetable. #1}
              \end{center}}
        \fi}

\def\@citex[#1]#2{\if@filesw\immediate\write\@auxout
	{\string\citation{#2}}\fi
\def\@citea{}\@cite{\@for\@citeb:=#2\do
	{\@citea\def\@citea{,}\@ifundefined
	{b@\@citeb}{{\bf ?}\@warning
	{Citation `\@citeb' on page \thepage \space undefined}}
	{\csname b@\@citeb\endcsname}}}{#1}}

\newif\if@cghi
\def\cite{\@cghitrue\@ifnextchar [{\@tempswatrue
	\@citex}{\@tempswafalse\@citex[]}}
\def\citelow{\@cghifalse\@ifnextchar [{\@tempswatrue
	\@citex}{\@tempswafalse\@citex[]}}
\def\@cite#1#2{{$\null^{#1}$\if@tempswa\typeout
	{IJCGA warning: optional citation argument
	ignored: `#2'} \fi}}

\def\fnt#1#2{\footnotetext{\kern-.3em
	{$^{\mbox{\sevenrm #1}}$}{#2}}}

 1
 1
 1

\font\tenrm=cmr10

\font\ninerm=cmr9

\begin{document}
\thispagestyle{empty}
{\baselineskip=12pt
\hskip 3.5in CALT-68-1965

\hskip 3.5in DOE RESEARCH AND

\hskip 3.5in DEVELOPMENT REPORT

\vspace{1.0cm}}
\centerline{\large \bf Evidence for Non-perturbative String
Symmetries\footnote{Work supported in part by the U.S. Dept. of Energy
under Grant No. DE-FG03-92-ER40701.}}
\bigskip
\bigskip
\centerline{\large John H. Schwarz}
\medskip
\centerline{\it California Institute of Technology, Pasadena, CA 91125, USA}
\bigskip
\bigskip
\bigskip
\centerline{{\it Presented at the International Congress of Mathematical
Physics}}
\centerline{{\it Satellite Conference ``Topology, Strings, and Integrable
Models''}}
\centerline{{\it Paris, July 1994}}
\vskip 2.8 truein
\parindent=1 cm

\begin{abstract}  String theory appears to admit a group of
discrete field transformations
-- called $S$ dualities -- as exact non-perturbative quantum symmetries.
Mathematically, they are rather analogous to the
better-known $T$ duality symmetries,
which hold perturbatively.  In this talk the
evidence for $S$ duality is reviewed and some speculations are presented.
\end{abstract}
\vfil\eject
\setcounter{page}{1}
\section{Introduction}

Non-compact global symmetries are a pervasive feature of supergravity theories.
Typically, the group $G$ is realized nonlinearly on scalar fields that
parametrize the
homogeneous space $G/H$, where $H$ is the maximal compact subgroup of
$G$.  The first example of this phenomenon, with $G = SL(2,\RR)$ and $H =
U(1)$,
was uncovered in 1976 in a version of $N = 4,\, D = 4$ supergravity by Cremmer,
Ferrara,  and Scherk.\cite{bib:cremmer1}  Curiously, this particular example
corresponds precisely to the low-energy effective field theory associated with
the most studied example of $S$ duality in string theory -- the toroidally
compactified heterotic string.  An analogous non-compact $E_7$ symmetry was
found in $N = 8,\, D = 4$ supergravity by Cremmer and Julia in
1978,\cite{bib:cremmer2}  and many other examples were worked out
thereafter.\cite{bib:salam}

In 1990, Font Iba\~nez, L\"ust, and Quevedo proposed\cite{bib:font} that an
$SL(2,\ZZ)$ subgroup of the $SL(2,\RR)$ of Ref. 1 should be an exact symmetry
of the
heterotic string toroidally compactified (\`a la Narain\cite{bib:narain}) to
four dimensions.  This discrete symmetry group is called $S$ duality, because
the $N=1$ superfield (containing the axion and dilaton) that parametrizes
$SL(2,\RR)/U(1)$ is often called $S$.  That $S$ duality should be an exact
symmetry
was a bold conjecture, since it implies, as a special case, an
electric-magnetic duality in which the coupling constant is inverted
($g_{el} \rightarrow g_{mag} \propto
1/g_{el}$).  Thus, if true, it has implications beyond perturbation theory.  On
the other hand, string theory is well understood only in perturbation theory.
A non-perturbative formulation is still lacking.  Thus, when it first appeared,
the proposal of Font {\it et al.},  seemed to me
(and probably to others, too) to be
intriguing but impossible to test.  As I will describe, this is not the
case.  Impressive non-trivial tests of $S$ duality have been formulated and
verified.  Thus my point of view now is that $S$ duality and its extensions
should be explored as broadly as possible as part of a program that may someday
culminate in a non-perturbative formulation of quantum string theory.
Indeed, the symmetry might well play a central role in the construction.

In string theory, as in classical supergravity, $S$ duality is realized by
field transformations.  As such, it is a statement about the theory and not
about classical solutions or quantum ground states. Any particular
solution will generically result in complete spontaneous breaking of the
symmetry.  In special cases (corresponding to orbifold points of the moduli
space) a small subgroup, such as  $\ZZ_2$ or $\ZZ_3$, may remain unbroken.
Thus
$S$ duality is analogous to a broken gauge symmetry.  Though it is
a discrete symmetry group, it is believed
(mostly by analogy with $T$ duality) to be a local symmetry of string
theory in the sense that field configurations that are  related by group
transformations should be identified and counted just once in the path
integral.  Note that this definition does not imply that group elements are
functions of space-time.  This is just as well since the fundamental string
action -- whatever it  may be -- is unlikely to make explicit reference to
space-time.  The existence of a space-time should be only a  (large-distance)
feature of a certain class of solutions.

For a variety of reasons, I suspect that the
$SL(2,\ZZ)$ group is a small piece of a much larger symmetry group.  The
complete
group might be the hyperbolic Kac--Moody group $E_{10}$,\cite{bib:gebert}  or
a discrete subgroup $E_{10} (\ZZ)$,\cite{bib:hull} or some continuous group
or supergroup containing $E_{10} (\ZZ)$. (Recall that $T$ dualities are special
cases of continuous group transformations.)

Another approach to uncovering the underlying symmetry in string theory,
pioneered by Gross and Mende,\cite{bib:gross} is the study of string amplitudes
at very high energy.  A more algebraic approach to obtaining much of
the same information has been developed recently
by Moore.\cite{bib:moore}  The
correspondence between the symmetries found by such methods and the $S$  and
$T$ duality symmetries that I will discuss, is not fully understood.  Since the
analyses of Refs. 8 and 9 are based on perturbation theory, they would seem
capable of revealing
$T$ duality symmetries only.  However, it may be possible to extract more
information than that. Shenker\cite{bib:sherker} has presented evidence that
string theory has
non-perturbative phenomena that are of form $\exp [- c/g]$ in addition to the
familiar instanton-like effects that go as $\exp [- c/g^2]$.  It would be very
interesting to see how this meshes with $S$ duality.  Another related issue
that needs to be better understood is the structure of the high-temperature
phase of string theory -- above the Hagedorn transition.\cite{bib:atick}  This,
too, requires non-perturbative information.  The answer is apparently essential
for an understanding of how string theory overcomes the much-discussed entropy
and information-loss problems associated with black holes.\cite{bib:susskind}

\section{Two Kinds of Duality}

$T$ duality or ``target space duality'' is a discrete symmetry of various
string theories that holds order by order in perturbation theory.  (For a
review see Ref. 13.)  In the simplest case of compactification on a circle the
group is $Z_2$ and the transformation corresponds to inversion of the radius
$(R \rightarrow \alpha'/R)$.  Of course, $R$ is determined by the value of
scalar field (a $T$ modulus) and the symmetry is realized as a field
transformation.  For any value other than $R = \sqrt{\alpha'}$, the symmetry is
spontaneously broken.  The generalization to toroidal compactification of the
heterotic string (\`a la Narain) has been studied in detail by many authors.
In this case no supersymmetry is broken and (for $d = 4$) there are 132 scalar
fields that live on the homogeneous space $M_0 = O(6,22) /O (6) \times  O
(22).$  However, the $T$ duality group is $G_T = O(6, 22; \ZZ)$ and the moduli
space is, therefore, $M = M_0/G_T$.  The 132 scalar fields belong to 22
abelian $N = 4$ gauge multiplets.

The toroidally compactified heterotic string also contains two additional
scalar fields -- called the axion $\chi$ and the dilaton $\phi$ -- which belong
to the $N = 4$ supergravity multiplet.  This supergravity theory is precisely
the one studied in Ref. 1, where $\chi$ and $\phi$ were shown to parametrize
$SL
(2, \RR)/U(1)$.  To show how this works, let us introduce a complex scalar
field.
\begin{equation}
\lambda = \chi + i e^{-\phi} = \lambda_1 + i \lambda_2 ~.
\end{equation}
In terms of $\lambda$, the symmetry is realized by linear fractional
transformations
\begin{equation}
\lambda \rightarrow {a \lambda + b\over c\lambda + d} ~,
\end{equation}
where $\left(\matrix{a & b\cr c & d\cr}\right)\, \in \, SL(2,  \RR)$.  The
classical value of $\lambda$ is $<\lambda > = {\theta\over 2\pi} + {8\pi i\over
g^2}$, where $\theta$ is the vacuum angle and $g$ is the coupling
constant.\footnote{Recall that $N=4$ Yang--Mills theories have vanishing
$\beta$ function, so that $\theta$ and $g^2$ are well-defined independent of
scale.    This feature presumably holds for heterotic strings in $N = 4$
symmetric backgrounds, as well.}  When instanton effects are taken into account
the Peccei-Quinn symmetry $\chi \rightarrow \chi + b$, which corresponds to the
$SL(2,\RR)$ subgroup given by matrices $\left(\matrix{1 & b\cr 0 &
1\cr}\right)$,
is broken to the discrete subgroup for which $b$ is an integer.  This subgroup
and the inversion $\lambda \rightarrow - 1/\lambda$ generate the discrete group
$SL(2,\ZZ)$ or $PSL(2, \ZZ)$.

Mathematically, $S$ and $T$ duality are quite analogous in the 4D low-energy
effective field theory (EFT) even though their implications for string theory
are dramatically different.  This analogy was one of the original motivations
for proposing that $S$ duality should also be a symmetry.  Let me briefly
describe the bosonic sector of the EFT and mention how the symmetries are
realized. (For more details see Ref. 14.)   The massless bosonic fields are the
metric tensor $g_{\mu\nu}$, the axion-dilaton field $\lambda$, (where $\chi$ is
related to the antisymmetric tensor field $B_{\mu\nu}$ by a duality
transformation), 28 abelian gauge fields $A_\mu^a$, and 132 moduli $M^{ab}$
parametrizing $O(6,22)/O (6) \times O(22)$.  These are the only massless fields
at generic points in the classical moduli space.  At the singular points,
which we ignore here, there are more.  The moduli $M^{ab}$ are conveniently
described as a symmetric $28 \times 28$ matrix belonging to the group O(6,22):
\begin{equation}
M^T = M, ~~ M^T LM = L \end{equation}
\begin{equation}
 L = \left( \matrix{0 & I_6 & 0 \cr I_6 & 0 & 0 \cr 0 & 0 & I_{22}}\right)
.\end{equation}

The classical action with the fields given above is
\[S = (const.) \int d^4 x \sqrt{-g} \Bigg( R - {g^{\mu\nu}\over 2
\lambda_2^2} \partial_\mu \lambda \partial_\nu \bar{\lambda}\]
\begin{equation}
 + {1\over 8} g^{\mu\nu} tr (\partial_\mu M L \partial_\nu ML) +
L_{gauge}\Bigg)
{}~,\end{equation}
where
\begin{equation} L_{gauge} = {\lambda_1\over 4} F_{\mu\nu}^a L_{ab}
\tilde{F}^{b\mu\nu} - {\lambda_2\over 4} (LML)_{ab}  F_{\mu\nu}^a F^{\mu\nu b}
{}~.\end{equation}
As usual, $F_{\mu\nu}^a = \partial_\mu A_\nu^a -\partial_\nu A_\mu^a$ and
$\tilde{F}^{\mu\nu} = {1\over 2\sqrt{-g}} \epsilon^{\mu\nu\rho\lambda}
F_{\rho\lambda}$.  Under $T$ duality, the transformation rules are
\begin{equation}
 M \rightarrow \Omega M \Omega^T, A_\mu \rightarrow \Omega A_\mu
{}~,\end{equation}
where $\Omega^T L \Omega = L$ and $g_{\mu\nu}$ and $\lambda$ are invariant.
Each term in the action is manifestly invariant under these transformations.
The group here is $O(6,22)$, but in string theory when one includes the
Kaluza--Klein and winding-mode excitations it is broken to the discrete
subgroup
$G_T = O(6,22; \ZZ)$.

Under $S$ duality one has
\begin{equation}
\lambda \rightarrow {a\lambda + b\over c \lambda + d}\end{equation}
\begin{equation}
 F_{\mu\nu}^a \rightarrow (c \lambda_1 + d) F_{\mu\nu}^a + c \lambda_2
(ML)_{ab} \tilde{F}_{\mu\nu}^b ~,\end{equation}
while $g_{\mu\nu}$ and $M^{ab}$ are invariant.  This is a symmetry of the first
three terms of the action, but not of $L_{gauge}$.  However, one can show that
it is a symmetry of the equations of motion that follow from $S$.  This
suggests there may be a different action giving rise to the same equations of
motion that has $S$ duality symmetry.  Such an action was constructed by Sen
and
me,\cite{bib:schwarz} but I will not present it here.
I will simply remark that we could
make $S$ and $T$ duality simultaneously manifest by sacrificing manifest
general coordinate invariance.  It is not known whether it is possible to
introduce auxiliary fields so as to realize all three symmetries off-shell at
the same time.  The problem is analogous to that for supersymmetry for which
auxiliary fields can sometimes be found and sometimes not.  If one insists on
off-shell closure of the algebra, it would be natural to include supersymmetry
at the same time.  Of course, our real interest is string theory, for which
off-shell issues still seem remote (and maybe even irrelevant).

\section{The Soliton Spectrum}

The 28 U(1) gauge fields $A_\mu^a$ give rise to 28 sets of electric and
magnetic charges.  A convenient way to define them is to assume that space-time
is asymptotically flat and use the asymptotic behavior of the
field strengths:
\begin{equation}
 F_{0i}^a \sim {q_{el}^a\over r^3} x^i~~\quad
\tilde{F}_{0i}^a \sim {q_{mag}^a\over r^3} x^i ~.\end{equation}
As explained in Ref. {14}, the allowed charges are then controlled by the
asymptotic values of the moduli ($\lambda \sim \lambda^{(0)}$ and $M_{ab} \sim
M_{ab}^{(0)}$) and a pair of vectors $\alpha_0^a, \beta_0^a$ belonging to the
Narain lattice, which is an even self-dual Lorentzian lattice of signature
(6,22).  The formulas are
\begin{equation}
 q_{el}^a = {1\over \lambda_2^{(0)}} M_{ab}^{(0)} (\alpha_0^b + \lambda_1^{(0)}
\beta_0^b), \quad q^a_{mag} = L_{ab}\beta_0^b ~.\end{equation}
A nice feature of these formulas is that they automatically incorporate the
Dirac--Schwinger--Zwanziger--Witten quantization rules ({\it i.e.}, the
quantization
condition for dyons with a $\theta$ angle).

The central charges that appear in the $N = 4$ supersymmetry algebra are
determined in terms of the electric and magnetic charges.  Moreover, the mass
of
a state is bounded below by its central charges.  This bound, known as the
Bogomol'nyi bound implies that an $N = 4$ multiplet with charges given by
$(\alpha_0^a, \beta_0^a)$ must satisfy
\begin{equation}
 ({\rm Mass})^2 \geq {1\over 16} (M^{(0)} + L)_{ab} (\alpha_0^a \, \beta_0^a)
{\cal M}^{(0)} \left(\matrix{\alpha_0^b\cr \beta_0^b}\right) ~,\end{equation}
where the $2\times 2$ matrix ${\cal M}^{(0)}$ is the asymptotic value of
\begin{equation}
{\cal M} = {1\over\lambda_2} \left(\matrix{1 & \lambda_1\cr
\lambda_1 & |\lambda|^2}\right) ~.
\end{equation}
This matrix transforms under $S$ duality in a manner analogous to the way $M$
transforms under $T$ duality.  Thus, the Bogomol'nyi bound is manifestly
invariant under both $S$ and $T$ duality provided that $(\alpha_0^a,
\beta_0^b)$ is a (2,28) representation of $G_S \times G_T$, which is indeed the
case.  Note that since $\alpha_0^a$ and $\beta_0^a$ belong to a lattice, it is
essential that these are the discrete groups $G_S = S L (2, \ZZ)$ and $G_T =
O(6,22; \ZZ)$.

This is very nice, but it certainly doesn't prove that $G_S \times G_T$ is  a
symmetry of the toroidally compactified heterotic string.  $T$ duality works
perturbatively and is well understood, but how can we prove $S$ duality without
knowing non-perturbative string theory?  As yet, we cannot prove it, but we can
subject the conjecture to some non-trivial tests.  Specifically, if we focus on
states that saturate the bound in eq. (12),
then we have ``short'' representations of the
$N = 4$ algebra.  Generically, a representation would have dimension $2^8 =
256$, but when the bound is saturated it has dimension $2^4 = 16$.
(The gauge multiplets with
vanishing masses and charges are examples.)  The analysis is analogous to that
for the Poincar\'e group, which admits short representations for massless
fields.
 The essential fact, pointed out long ago by Witten and Olive,\cite{bib:witten}
is that such states can receive no quantum corrections -- perturbative or
nonperturbative -- to their masses so long as the supersymmetry remains
unbroken.  Accepting that, we need only establish that the  degeneracies of
such 16-dimensional representations, $N_{16} (\vec{\alpha}_0, \vec{\beta}_0)$,
are $SL(2, \ZZ)$ invariant.  Note that because the vacuum breaks $S$ and $T$
duality spontaneously, the states saturating the Bogomol'nyi bound do
not form degenerate multiplets of these groups.  Eq. (12) is invariant only
when ${\cal M}^{(0)}$ and $M^{(0)}$ are transformed together with $(\alpha_0^a,
\beta_0^a)$.  If specific values of ${\cal M}^{(0)}$ and $M^{(0)}$ leave a
subgroup of $G_S \times G_T$ unbroken, then degenerate multiplets correspond to
representations of that subgroup.

Let us now explore which states in the elementary string spectrum saturate the
Bogomol'nyi bound.  First of all, such states carry electric charges only
(associated with internal momentum and winding-mode excitations).  Thus,
absorbing $M^{(0)}$ in the definition of the lattice, eq. (12) simplifies to
\begin{equation}
({\rm Mass})^2 = {1\over 16 \lambda_2^{(0)}} \hat{\alpha}^a
(I + L)_{ab} \hat{\alpha}^b =
{1\over 8\lambda_2^{(0)}} (\hat{\alpha}_R)^2 ~,\end{equation}
where $\hat{\alpha} \cdot \hat{\alpha} = \hat{\alpha}_L \cdot \hat{\alpha}_L -
\hat{\alpha}_R \cdot \hat{\alpha}_R$.  ($\hat{\alpha}_L$ is 22-dimensional and
$\hat{\alpha}_R$ is 6-dimensional.  They correspond to the left-moving and
right-moving internal momenta of the string.)  Now we should compare the  free
string spectrum, which is given by
\begin{equation}
({\rm Mass})^2 = {1\over 4\lambda_2^{(0)}} \left[ {1\over 2}
(\hat{\alpha}_L)^2 + N_L - 1\right]
= {1\over 4\lambda_2^{(0)}} \left[ {1\over 2} (\hat{\alpha}_R)^2 + N_R -
\delta\right] ~.
\end{equation}
$N_L$ and $N_R$ represent left-moving and right-moving oscillator excitations.
The parameter $\delta$ is 1/2 in the NS sector and $0$ in the R sector.
Alternatively, it is simply 0 in the $GS$
formulation. The factor of $(\lambda_2^{(0)})^{-1}$ appears because the mass is
computed with respect to the canonically normalized Einstein  metric.  It does
not appear if one uses the string metric, which differs by a dilaton-dependent
Weyl rescaling.  The Einstein metric is more natural in the present context,
because it is invariant under $S$ duality transformations.  Comparing formulas,
one sees that the Bogomol'nyi bound is saturated provided that $N_R =
\delta$ (which gives 8 bosonic and 8 fermionic right-moving modes -- the short
representation of $N = 4)$ and $N_L = 1 + {1\over 2} \hat{\alpha} \cdot
\hat{\alpha}$.  Thus, if $\hat{\alpha} \cdot \hat{\alpha} = 2n - 2$, for a
non-negative integer $n$, then there is a short $N = 4$ multiplet for every
solution of $N_L = n$.  These states are only ``electrically'' charged.  The
challenge is to find their predicted $S$ duality partners.  Specifically, every
elementary string excitation of the type we have just described $(\vec{\alpha}
= \vec{\ell}, \vec{\beta} = 0)$ should have magnetically charged partners with
$\vec{\alpha} = a \vec{\ell}$ and $\vec{\beta} = c \vec{\ell}$. Since $a$ and
$c$ are elements of an $SL(2,\ZZ)$ matrix, they are relatively prime
integers.  If such states exist, one can use
the Bogomol'nyi bound to prove that
they are absolutely stable.  This is important since the formula for
the mass, which
is supposed to be exact, is real.

Sen has investigated the partners of electrically charged states with
$\vec{\ell} \cdot \vec{\ell} = - 2 \,(N_L = 0)$.  He has explained that
$S$ duality partners with $c = 1$
can be identified with BPS monopole solutions
(and their dyonic generalizations) of the EFT.
These solutions saturate the bound, of
course.  Thus, as we have explained, they should persist with exactly this mass
in the complete quantum
string theory.  For $c > 1$, Sen argues that one should examine
multi-BPS dyon bound states.  Specifically, he shows that
the prediction of $S$ duality is
that each multi-BPS dyon moduli space should admit a unique normalizable
harmonic
form.  Poincar\'e duality would give a second one unless it is self-dual or
anti-self-dual.  He then constructs such an anit-self-dual
form explicitly for the case of $c =
2$,\cite{bib:sen1} providing the most non-trivial test of $S$ duality yet.
Progress toward extending this result to $c > 2$ is attributed to G. Segal
(unpublished).

The next case to consider is $\vec{\ell} \cdot \vec{\ell} = 0$, corresponding
to $N_L = 1$.  This case probes stringy structure, which makes it
especially interesting.
Since states with $N_L = 1$ are given by a single oscillator
excitation $\alpha_{-1}^\mu | 0 >$, one should find states with helicity
$\pm 1$ once each
and ones with helicity $0$ 22 times. Solitons with
$\vec{\ell} \cdot \vec{\ell} = 0$ have been obtained\cite{bib:harvey} by
wrapping five-branes around the torus.  They are called ``$H$ monopoles''
because the gauge fields in question arise from dimensional reduction of $H =
dB$ (rather than the metric or 10-dimensional gauge fields).  The question then
is again the cohomology of the corresponding moduli space.  The relevant part
of the moduli space $\hat{\cal M }$ is four-dimensional and hyper-K\"ahler.  A
normalizable harmonic $(p,q)$ form would give a state with left-moving
helicity ${1\over 2} (p- q)$.  In their first paper  on the subject,
Gauntlett and Harvey noted that K3 is a hyper-K\"ahler space with precisely the
desired cohomology. However, they soon realized that it does not have the
required $SO(3)$ isometry, and so they suggested\cite{bib:gauntlett} that
actually $\hat{\cal M } = R^4/\ZZ_2$. The cohomology of this space
only gives 8 of the 24 desired states, so they
suggested that the 16 remaining spin-0 states might arise from a stringy
twisted sector. This example still needs more study. In particular, the
noncompactness of $\hat{\cal M }$ gives rise to infra-red issues that need to
be clarified.

\section{Concluding Discussion}

We have shown that $S$ duality is a classical symmetry of $N = 4, D = 4$
supergravity coupled to abelian gauge multiplets.  The hypothesis
that it is an exact symmetry of string theory leads to specific testable
predictions for the soliton spectrum of the heterotic string compactified on a
torus.  Several magnetically charged states have been shown to occur with
precisely the predicted properties.  Because of the essential way in which $N =
4$ supersymmetry and the Bogomol'nyi bound are used,  it was important
to carry out these tests for a special class of vacua (ones with unbroken $N =
4$ supersymmetry and abelian unbroken gauge symmetries).  However, the
underlying symmetry is realized in terms of field transformations, and so the
particular choice of vacuum shouldn't be of fundamental importance, just a
matter of convenience.

Recently, Girardello {\it et al.} and
Vafa and Witten have examined the implications of $S$ duality for $N
= 4$ Yang--Mills,\cite{bib:vafa} and Seiberg and Witten have examined $N = 2$
Yang--Mills.\cite{bib:seiberg} I will not review those beautiful works here,
but simply make a few remarks.  First of all, in these theories the field
$\lambda$ is replaced by its value $<\lambda> = {\theta\over 2\pi} + i
{8\pi\over g^2}$.  Therefore, $S$ duality is no longer a symmetry realized by
field transformations, but rather a rule stating that the theory for  one value
$<\lambda>$ is equivalent to the one for a transformed value -- provided one
suitably redefines the electric and magnetic charges at the same time.  Such a
duality was conjectured long ago by Montonen and Olive.\cite{bib:mortoren}  In
the $N=4$ case, Vafa and Witten show that the partition function $Z(\lambda)$
transforms under $SL(2,\ZZ)$ as a modular form whose weight is determined by
the
Euler characteristic of the four-manifold on which the theory is defined.  (In
the full string theory, I would expect other fields to give a compensating
contribution so that the $\lambda$ integral is modular invariant.)  The $N = 2$
case is more subtle since $<\lambda>$ is scale dependent
when the theory is asymptotically
free.  The appropriate object to compute turns out  to be the quantum moduli
space, which is described by an elliptic (or hyperelliptic\cite{bib:klemm})
curve.  The reader is referred to Ref. [21] for more information.

In conclusion, our understanding of $S$ duality is progressing rapidly.  It
should be helpful in the search for
a more fundamental formulation of string theory.
Also, its implications for vacua with $N = 1$ supersymmetry could be of
phenomenological interest.

\end{document}